# Strong phonon softening without Fermi surface nesting

D. Lamago[1,2], M. Hoesch[3], M. Krisch[3], R. Heid[1], K.-P. Bohnen[1], P. Böni[4] and D. Reznik[1]
[1]*Forschungszentrum Karlsruhe, Institut für Festkörperphysik, P.O.B. 3640, D-76021 Karlsruhe, Germany*
[2]*Laboratoire Léon Brillouin, CEA Saclay, F-91191 Gif-sur-Yvette, France*
[3]*European Synchrotron Radiation Facility,F-38043 Grenoble Cedex, France*
[4]*Physik Department E21, Technische Universität München, D-85748 Garching, Germany*

Interactions with electronic excitations can soften and/or broaden phonons. They are greatly amplified at wavevectors $\mathbf{Q}^{\pm}$ that connect parallel (nested) sheets of the Fermi surface. In such a case, called a Kohn anomaly, the phonon dispersion sharply dips and its linewidth has a maximum at $\mathbf{Q}^{\pm}$. Here we present results of inelastic x-ray scattering measurements that uncovered soft phonons in chromium far from $\mathbf{Q}^{\pm}$. They appear in addition to the previously reported soft phonons at $\mathbf{Q}^{\pm}$. Calculations in the local density approximation (LDA) show that the new anomalies originate from enhanced electron-phonon coupling. A similar mechanism may explain phonon anomalies away from nesting wavevectors in copper oxide superconductors and other compounds.

**PACS numbers:** 74.25.Kc, 78.70.Nx, 63.20.-e, 71.38.-k
**Keywords**: Spin density wave, inelastic X-ray scattering, Kohn anomalies, Electron-phonon coupling

Incommensurability is at the heart of many interesting properties of a large class of materials. For example, in the parent compound $La_2CuO_4$, doping leads to the formation of incommensurate spin fluctuations that may promote high temperature superconductivity. [1] The formation of incommensurate cycloidal spin structures in mulitferroic materials has been identified as a possible coupling mechanism of ferroelectric with magnetic and lattice degrees of freedom. [2] In a third class of systems, like MnSi, the appearance of incommensurability is intimately related to the lack of a center of symmetry of the crystal structure. It leads via the spin-orbit interaction to the formation of incommensurate helical order due to the competition between the Dzyaloshinsky-Moriya (DM) interaction and the exchange interaction [3]. Application of pressure leads to partial magnetic order and non-Fermi liquid behavior [4]. In the cuprates, competing explanations of incommensurability are either in terms of stripe order or of charge and spin density waves caused by nesting of the Fermi surface. If nesting occurs phonon broadening and/or softening at the nesting wavevectors may be expected. Thus previous observations of strong phonon anomalies in cuprate superconductors [5] may be interpreted as evidence for Fermi surface nesting.

Cr is one of the most famous materials exhibiting incommensurate order. Although Cr crystallizes in a simple bcc-structure, its Fermi-surface exhibits a rather complicated shape. As first pointed out by Lomer [6], a spin-density wave occurs due to the nesting of the "jack-shaped" electron Fermi surface at the Γ-point (0,0,0) with the octahedral shaped hole surface at the H-point (1,0,0). At the Néel temperature $T_N$ = 311 K, Cr undergoes a transition to a transversely polarized spin-density-wave (TSDW) at an incommensurate wave vector $\mathbf{Q}^{\pm}=(1\pm\delta,0,0)$ where $\delta \approx 0.048$. At $T_{sf}$ = 121 K, a first order phase transition takes place from the TSDW phase to a longitudinally polarized spin-density-wave (LSDW) phase with the same modulation vector, where the moments are oriented along $\mathbf{Q}^{\pm}$.

The magnetic excitation spectrum in Cr also exhibits many unusual features. In the LSDW phase, incommensurate excitations with longitudinal and transverse polarization with respect to the staggered magnetization emerge from the magnetic satellite peaks with an extremely steep dispersion of the order of 1000 meVÅ. In addition, diffuse scattering evolves in the region between the satellite positions and becomes stronger with increasing temperature and energy. Most remarkably, additional low-energy modes with longitudinal polarization appear (Fincher-Burke modes) above $T_{sf}$ in the TSDW-phase. While the incommensurate excitations can be essentially explained in terms of the three band model [7], the nature of the Fincher-Burke modes is not yet understood [8,9].

Associated with the SDW, there is a distortion of the lattice with twice the SDW wave vector, $\mathbf{Q}^{\pm}$. Charge density waves (CDW) were observed there using synchrotron radiation [10] confirming previous results by x-ray [11] and neutron diffraction [12]. In addition higher harmonics of the SDW as well as the CDW were observed. The occurrence of the CDW can be either explained in terms of Fermi surface nesting [13] or by a strain wave induced by the magnetoelastic coupling to the SDW.

Interaction of the conduction electrons with the lattice vibrations enhanced by the Fermi surface anomalies such as nesting leads to anomalous phonon dispersions at wave vectors of extremal dimensions of the Fermi surface (Kohn anomalies). Using inelastic neutron scattering (INS), Shaw and Muhlestein [14] have identified four regions of anomalous transverse phonon softening in Cr, which they associated with the topology of the Fermi surface. The most pronounced ones occur near the H-point and the N-point. (see fig. 1a) The relatively poor wavevector ($\mathbf{Q}$) resolution, however, washed out the phonon anomalies, thus precluding a direct comparison of the anomalous wavevectors with the position of the magnetic peaks.

Our goal was to precisely determine the wavevectors of phonon anomalies in Cr and to determine their correspondence with the FS nesting as well as the CDW and SDW wavevectors. In order to answer this question, which is of general interest for materials with nested Fermi surfaces, we have investigated the phonon dispersion near the H-point and extended the measurements along the zone boundary to the N-point where another phonon anomaly occurs [14]. To improve the **Q**-resolution we used inelastic x-ray scattering (IXS). We furthermore took advantage of the capability of IXS instruments to easily change the scattering plane and to trace the phonon anomaly away from high symmetry directions.

Figure 1a shows results of our Born-von-Karman model calculation with parameters fitted to previously published neutron data represented by data points. It shows transverse (T) and longitudinal (L) acoustic phonon branches. There are two transverse acoustic branches in the [110] direction, T1 and T2. The strongest previously reported phonon anomalies appear as softening of the measured frequencies at positions marked by arrows below the calculated curve.

The IXS experiments were performed at the ESRF on the ID-28 beam line using (9 9 9) Si reflections, providing an energy resolution of 3 meV full width at half maximum. The sample was a high quality single crystal of Cr in the “multi-Q” state. It was mounted in a closed-cycle refrigerator with the temperatures varying between 50 K and 320 K. We measured phonon frequencies in the Brillouin zone centered at the **Q**=(3,1,0) reciprocal lattice position, where the structure factor of the anomalous phonons is maximized. The zone boundary was at **Q**=(3,0,0) in the [100] direction and at **Q**=(3,0.5,0.5) in the [110] direction. Moving between these two positions along the zone boundary, the [110] T2 branch connects to the [100] T branch (Fig. 1a). This made it possible to investigate the phonon anomaly along the entire 2-dimensional cut through the BZ without the restriction of the high symmetry direction. Figure 1b shows an example of raw data at **Q**=(0.5 3.5 0) corresponding to the [110] T2 branch and fitted with a Lorentzian. The blue lines in figure 2a show the positions in reciprocal space along which we measured the phonon frequencies.

The effects of electronic excitations across the Fermi surface on phonons can only be modeled by very long-range forces or extra attractively interacting shells if a shell model is used [15]. Thus, to isolate the effects of conduction electrons, we first compared the experimental dispersions with the Born-van-Karman model, which includes only short range interatomic forces.

Figure 2 compares calculated and measured phonon dispersions along the lines in reciprocal space shown in Figure 2a. Figure 2b,c shows the IXS results where previous neutron measurements were conducted. We observe clear dispersion minima at h=0.95 in the [1 0 0] direction and at h=0.5 in the [110] direction. These results convincingly demonstrate that the position of the minimum of the phonon dispersion corresponds to the magnetic satellite peaks of the SDW, not to the CDW positions. There was no change of the position or magnitude of the dip of the dispersion with temperature variation between 50K and 320K. In particular, the dip survives in the paramagnetic phase.

We then investigated the evolution of the phonon anomaly between **q**=(0.5,0.5,0) and **q**=(1,0,0) (Fig. 2d-g). The experimental results as compared with the predictions of the Born-van-Karman model show that the phonon dispersion minimum persists all along the zone boundary line connecting the two wave vectors (between the N and H points in figure 1a). Thus, anomalous phonon softening is not limited to the high symmetry directions but extends throughout the zone boundary.

Fig. 2h shows the difference between the Born-von-Karman model and the experimental values along the [100] direction (red) and along the zone boundary (blue). Along the [100] direction, the phonon softening has a distinct maximum at **q**=(0.95 0 0), the position of the SDW satellite as discussed above. The anomaly is also present along the entire zone boundary line between **q**=(0.5,0.5,0) and **q**=(1,0,0). The strongest renormalization in the anomalous region is 2 meV at **q**=(0.5,0.5,0) and the weakest is 1meV at **q**=(1,0,0). The line width of all phonons remains resolution-limited (≤ 0.5 meV) within the experimental uncertainty at all **q**.

To understand the mechanism behind these phonon anomalies we performed density functional calculations of the lattice dynamics using the mixed basis pseudopotential method and the linear response technique [16,17,18]. In the construction of a Vanderbilt-type pseudopotential [19] for Cr, 3s and 3p semicore states were included in the valence space. The fairly localized semicore states are treated efficiently in the mixed-basis scheme employing a combination of local functions and plane waves for the representation of the valence states. Local s, p, and d orbitals had a radial cutoff of 2.1 Bohr, supplemented by plane waves up to a kinetic energy of 24 Ry. For the exchange-correlation functional the local density approximation (LDA) in the parameterization of Perdew-Wang [20] was applied. We only considered non-magnetic states. The present results are obtained for the optimized lattice constant of 2.788 Å, which shows the typical overbinding of LDA with respect to the experimental value of 2.88 Å. Brillouin-zone (BZ) integrations were performed using 888 k points in the irreducible wedge of the BZ (IBZ) in connection with a smearing technique employing a Gaussian broadening of 0.2 eV. The complete phonon dispersion was obtained by standard Fourier interpolation of the dynamical matrices calculated at 145 **q** points in the IBZ. These parameters guaranteed sufficient convergence of phonon frequencies except in the anomalous regions. Additional calculations with 22960 k points in the IBZ and reduced broadenings of 0.1 and 0.05 eV have been performed for selected reduced momentum transfers **q** along [100] and [110].

The phonon self-energy depends on the electron-phonon coupling strength and density of states of electron-hole excitations. It is widely believed that the former does not strongly depend on the phonon wavevector, whereas the latter can have sharp features in reciprocal space if there is nesting of the Fermi surface. Thus features of the Fermi surface are usually held responsible for sharp dips in the phonon dispersion at specific wavevectors.

The density of states of electron-hole excitations at the Fermi surface that we calculated in the LDA is shown as a color map in Fig. 3. There are two features, the bright spots at $\mathbf{Q}^{\pm}$ away from the H-point and the weaker ones near the N-point. Both have been held responsible for the previously reported phonon softening around these wavevectors. [14] However a simple association between these nesting features and the phonon softening is not a proof of a cause and effect relationship. In fact phonon anomalies in the form of sharp dispersion dips may originate from either the Fermi surface nesting (Kohn anomalies) [21,22] or from the strong enhancement of the electron-phonon matrix elements at specific **q**-values [15,18].

Unlike the calculation based on the density of states of excitations across the Fermi surface, our linear response LDA phonon calculation includes not only the nesting effects but also the electron-phonon coupling strength. Here we propose a relatively simple method to reliably distinguish between the two types of anomalies. It is based on the idea that numerical smearing of the electronic bands in the LDA suppresses the Fermi surface nesting, but it should not have much impact if strong electron phonon coupling dominates the mechanism of the softening. Such smearing is always used for the numerical approximations to work, but here we employ it as a tool to distinguish between alternative mechanisms behind the phonon anomalies.

The calculation along the high symmetry directions (Fig. 4 a,b) was performed with two different values of Fermi surface smearing. The calculations away from the high symmetry directions were performed using a different numerical grid. All produced the same result at **q**=(0.5,0.5,0).

Figure 4 compares our results with the LDA-calculated phonon dispersions. The LDA gives the correct wavevectors of the strongest softening, but overestimates the strength of the effect, especially at **q**=(0.94,0,0) (Fig. 4b). There are no nesting features in the color plot (Fig. 3) corresponding to the softening along the line **q**=(h,1-h,0) (blue line in Fig. 2h). In contrast LDA indeed agrees reasonably well with the experiment (Fig. 4c,d). Thus the observed softening must be caused by enhanced electron-phonon coupling alone. We speculate that the appearance of such an anomaly at the zone boundary is not a coincidence, because the eigenvectors of these phonons have a higher symmetry leading to resonant behavior with certain electronic states. More work is necessary to prove or disprove this point.

The anomaly at $\mathbf{Q}^{\pm}$ is very sensitive to the smearing, whereas the one at the N-point is not. This observation confirms that only the $\mathbf{Q}^{\pm}$ anomaly originates primarily from the topology of the Fermi surface, i.e. nesting, whereas the others, result from the **q**-dependence of electron-phonon coupling. Both, the calculations and experiments agree that the maximum phonon softening near the H-point occurs at the wave vector of the SDW, not of the CDW. This result strongly favors the SDW-induced strain wave as opposed to Fermi surface nesting as the origin of the CDW. The calculation predicts a much stronger phonon softening than observed, perhaps because electronic correlations or magnetism not included in the LDA, wash out the Fermi surface. We note that a **Q**-resolution effect would greatly enhance the observed linewidth due to the averaging over a wide range of frequencies, but this was not observed.

An enhancement of the electron-phonon matrix elements has been proposed to explain a phonon anomaly in $NbSe_2$ [23,24] too. Therefore, our results may help in understanding a wide range of materials. They may also be relevant to the origin of the giant phonon anomalies half-way to the zone boundary in the high $T_c$ cuprates in the direction along the Cu-O bond. These are not reproduced by LDA [25], but the current work, as well as previous investigations [15,18], make it clear that the possibility of electron-phonon coupling enhanced by electronic correlations as opposed to a "hidden" Fermi surface nesting or collective modes should be seriously considered.

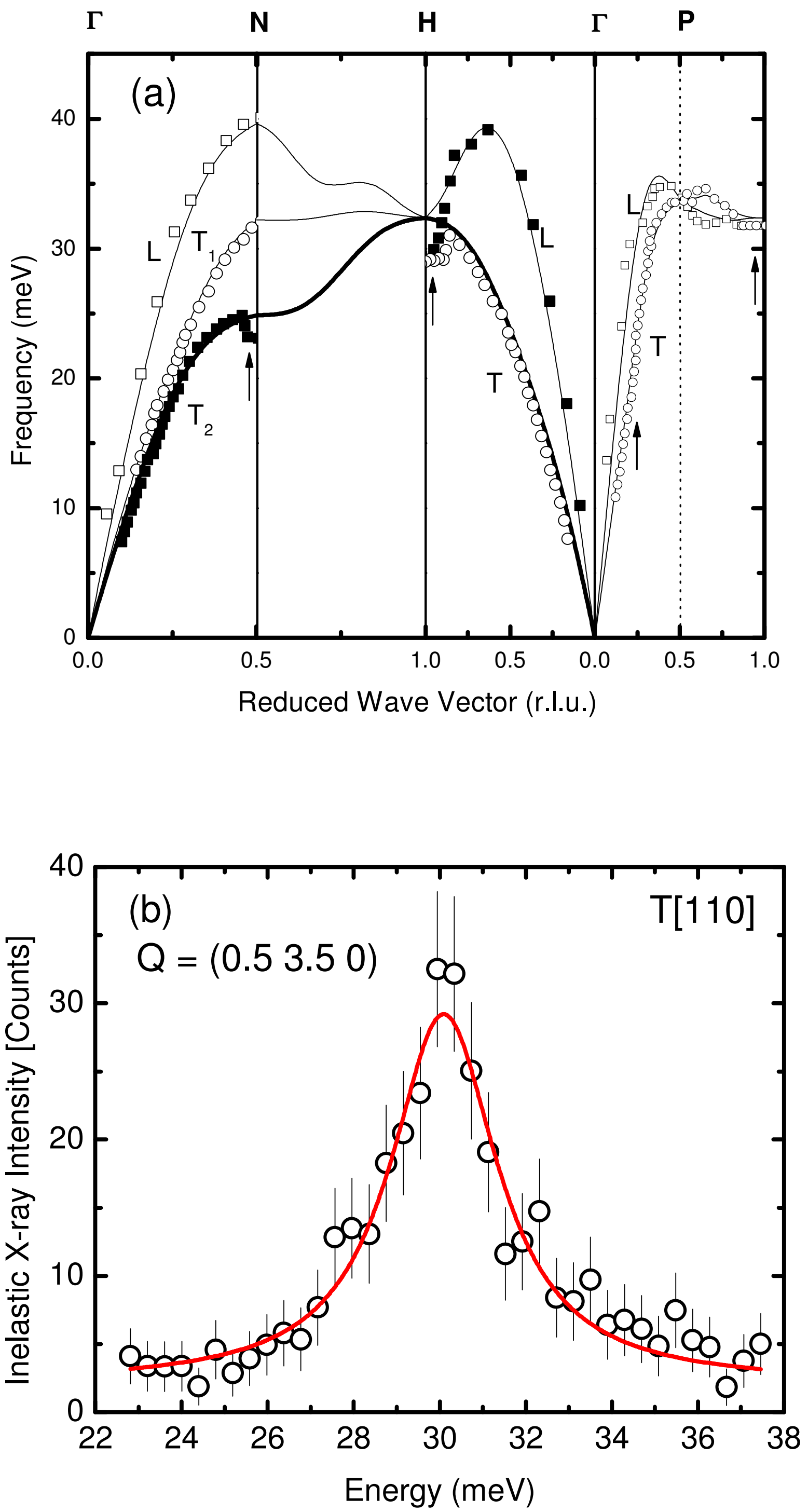

FIG. 1. (a) Phonon dispersion as calculated by means of the Born-van-Karman model along the high symmetry directions. Data points represent previous neutron data [9]. Arrows indicate anomalies as observed by neutron scattering [9]. Branches of interest here are shown in bold. (b) Raw data from inelastic X-ray scattering as function of energy transfer at T = 320 K and Q = (0.5 3.5 0). The solid line corresponds to a fit to the data using a Lorentzian.

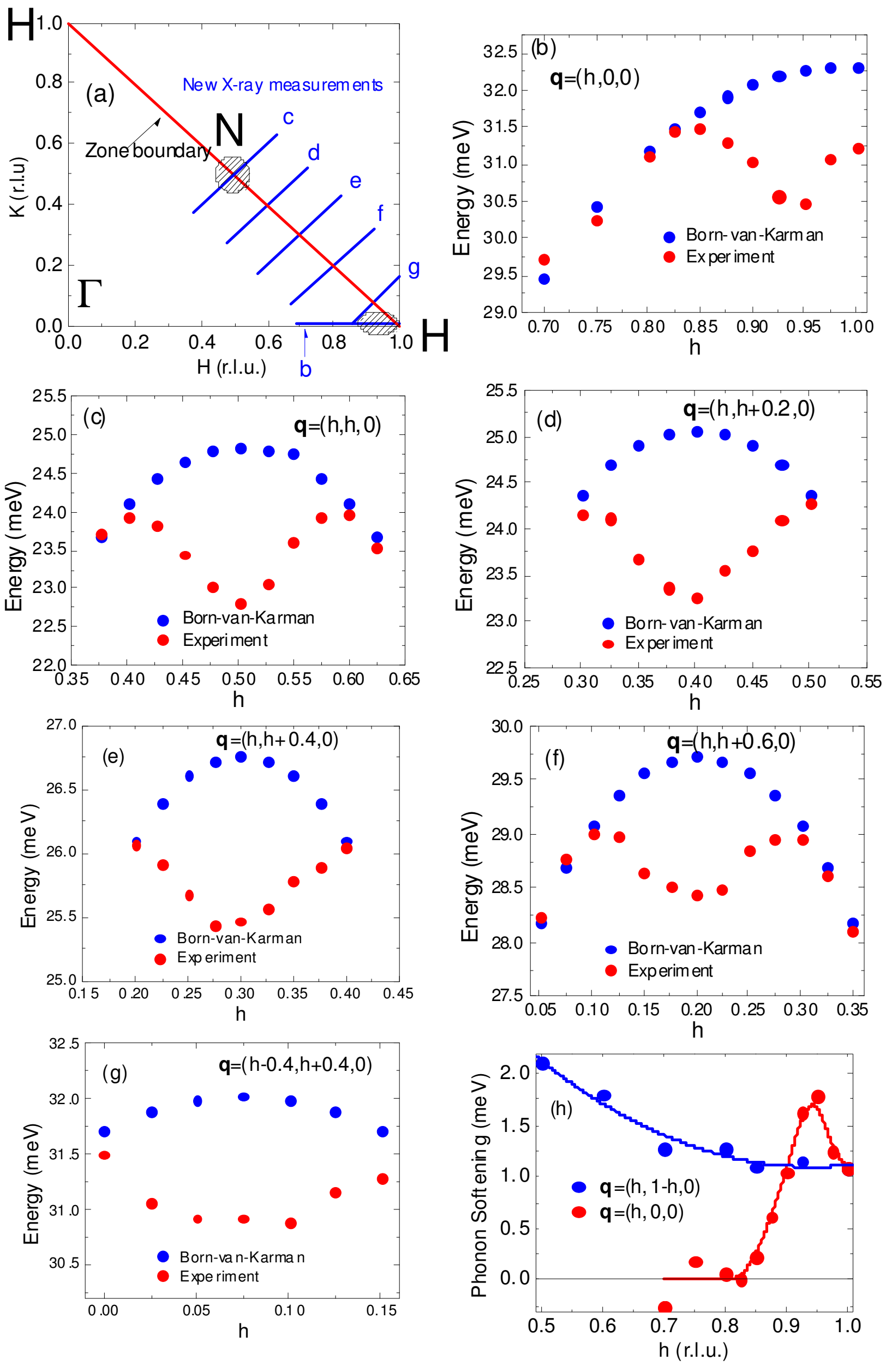

Figure 2. Comparison between experiment and Born-von-Karman calculations. The temperature was 300K. a) A schematic of the scans in reciprocal space. Hatched ovals indicate previously observed phonon anomalies; b-g) Calculated phonon dispersions along the paths indicated in (a); h) Difference ($E_{calc} - E_{exp}$) between the calculated and experimental phonon dispersion along the zone boundary N-H (blue) and along the high symmetry direction Γ-H (red).

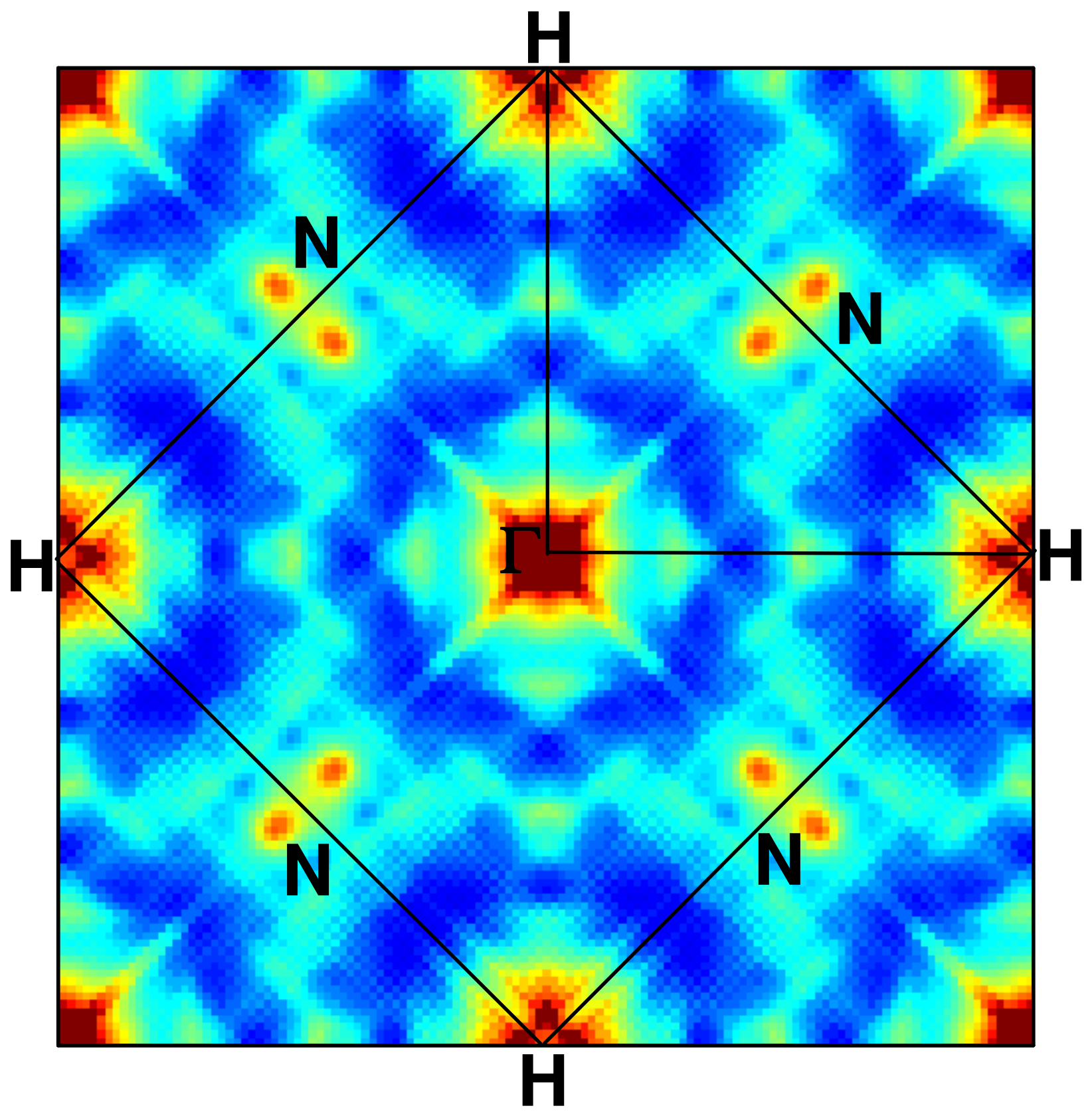

Figure 3: Joint density of states for electron-hole excitations, whose “hotspots” (in red) correspond to potential phonon anomalies.

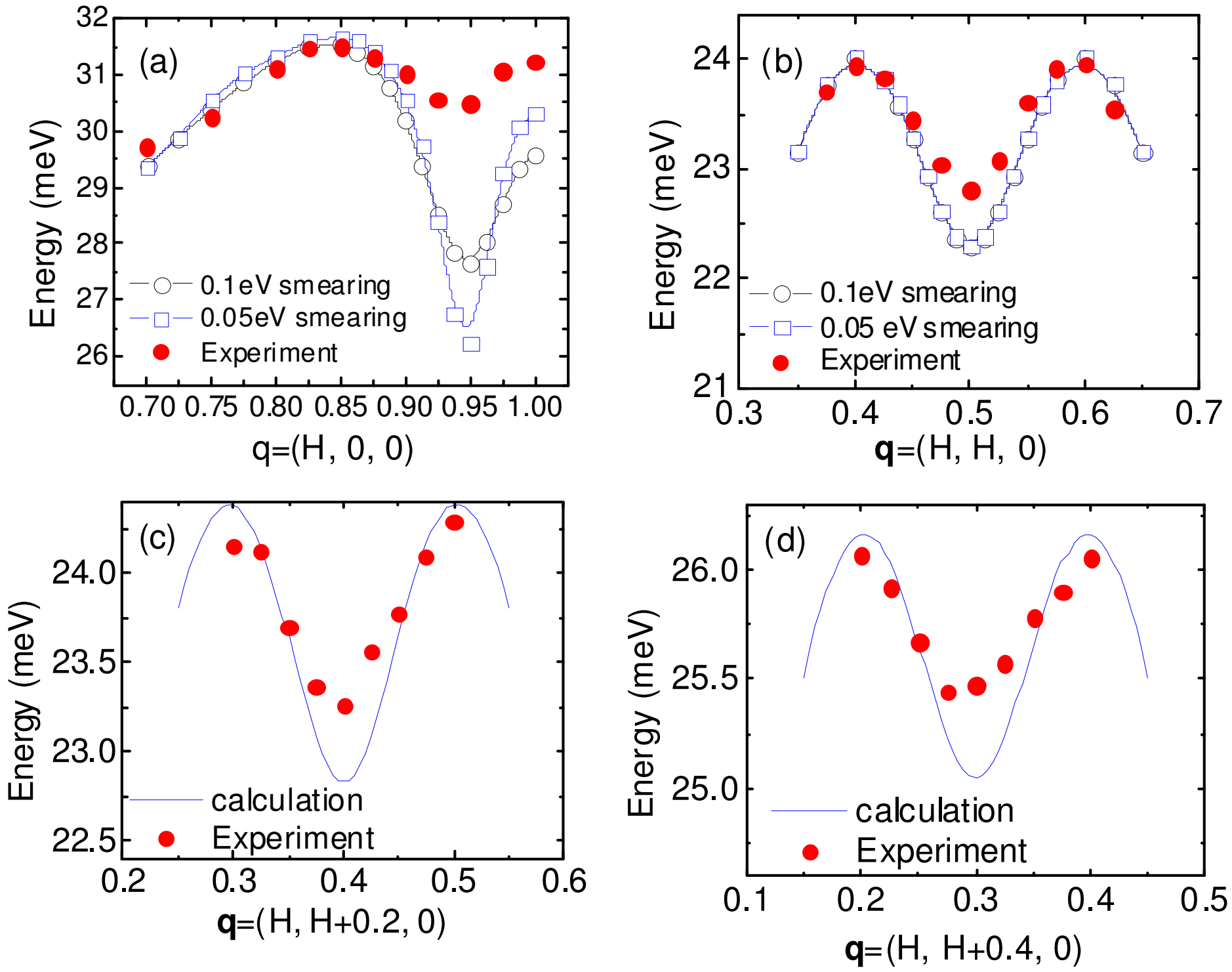

Figure 4. Comparison between phonon dispersions and the LDA calculation. (a,b) Black open circles represent a factor of 2 more smearing than the blue open squares. (c,d) Experiment and calculation away from the high symmetry directions.